\documentclass[journal]{IEEEtran}
\makeatletter
\long\def\@makecaption#1#2{\ifx\@captype\@IEEEtablestring%
	\footnotesize\begin{center}{\normalfont\footnotesize #1}\\
		{\normalfont\footnotesize\scshape #2}\end{center}%
	\@IEEEtablecaptionsepspace
	\else
	\@IEEEfigurecaptionsepspace
	\setbox\@tempboxa\hbox{\normalfont\footnotesize {#1.}~~ #2}%
	\ifdim \wd\@tempboxa >\hsize%
	\setbox\@tempboxa\hbox{\normalfont\footnotesize {#1.}~~ }%
	\parbox[t]{\hsize}{\normalfont\footnotesize \noindent\unhbox\@tempboxa#2}%
	\else
	\hbox to\hsize{\normalfont\footnotesize\hfil\box\@tempboxa\hfil}\fi\fi}
\makeatother

\usepackage{graphicx}
\usepackage{float}
\usepackage{wrapfig}
\usepackage{algpseudocode}
\usepackage{multirow}
\usepackage{amsfonts}
\usepackage{mathrsfs}

\usepackage{amssymb,latexsym}
\usepackage{commath}
\usepackage{caption}
\usepackage{subcaption}

\usepackage{booktabs}
\usepackage{mathtools}

\usepackage{todonotes}
\usepackage{soul}
\usepackage{algorithm}
\usepackage{algpseudocode}
\usepackage{algpseudocode} 
\usepackage{multicol}
\usepackage{amsbsy}

\usepackage[noadjust]{cite}

\usepackage[T1]{fontenc} % optional enhanced font encoding
\usepackage{amsmath}

\usepackage[cmintegrals]{newtxmath}
\usepackage{bm}

\usepackage{array}

\usepackage{url}

\usepackage[noadjust]{cite}

\ifCLASSINFOpdf

\else

\fi

\usepackage[T1]{fontenc} % optional enhanced font encoding

\usepackage[cmintegrals]{newtxmath}
\usepackage{bm}

\usepackage{array}

\usepackage{url}

\ifCLASSINFOpdf
\else
\fi
\providecommand{\hypersetup}[1]{\relax}

\hypersetup{pdftitle={Bare Demo of IEEE\_lsens.cls for IEEE Sensors Letters},%<!CHANGE!
pdfsubject={Typesetting},%<!CHANGE!
pdfauthor={Michael D. Shell},%<!CHANGE!
pdfkeywords={Class, IEEE, IEEE\_lsens, IEEE Sensors Letters, LaTeX, Typesetting, TeX}}%<^!CHANGE!}

\usepackage{color}

\graphicspath{{figs/}}

\begin{document}

%\title{PDA-Net: An Adversarial Domain Adaptative Neural Network for Denoising STM images}

\title{ Physics-augmented Deep Learning with Adversarial Domain Adaptation: Applications to STM Image Denoising}

\author{Jianxin Xie$^*$, Wonhee Ko, Rui-Xing Zhang, Bing Yao 
	
\thanks{
	Jianxin Xie ($^*$corresponding author: hcf7fd@virginia.edu) is with the School of Data Science, University of Virginia, Charlottesville, VA 22904 USA. Wonhee Ko and Rui-Xing Zhang are from the Department of Physics, University of Tennessee at Knoxville. Bing Yao (byao3@utk.edu) is with the Department of Industrial and Systems Engineering, University of Tennessee at Knoxville. 
}}

\maketitle

\begin{abstract}
Image denoising is a critical task in various scientific fields such as medical imaging and material characterization, where the accurate recovery of underlying structures from noisy data is essential. Although supervised denoising techniques have achieved significant advancements, they typically require large datasets of paired clean-noisy images for training. %, which are challenging to obtain. Insufficient paired datasets can lead to ineffective denoising models, failing to accurately capture the complexity of real-world noise. 
Unsupervised methods, while not reliant on paired data, typically necessitate a set of unpaired clean images for training, which are not always accessible.
%In quantum material research, scanning tunneling microscopy (STM) provides unparalleled insights into the electronic structure of materials at atomic resolution. However, real-world STM images are inevitably marred by noise, obscuring critical atomic details. The absence of clear benchmark datasets further exacerbates the difficulty of denoising real-world STM images. To address these challenges, 
In this paper, we propose a physics-augmented deep learning with adversarial domain adaption (PDA-Net) framework for unsupervised image denoising, with applications to denoise real-world scanning tunneling microscopy (STM) images. The underlying physics model is first used to generate simulation data and envision the ground truth for denoised STM images. %To deal with the limited sample size in the real domain, we created a supplementary dataset consisting of simulated blurry images for augmenting the existing real-world dataset and enhancing the model's ability to generalize across different image modalities. 
Additionally, built upon Generative Adversarial Networks (GANs), we incorporate a cycle-consistency module and a domain adversarial module into our PDA-Net to address the challenge of lacking paired training data and achieve information transfer between the simulated and real experimental domains. Finally, we propose to implement feature alignment and weight-sharing techniques to fully exploit the similarity between simulated and real experimental images, thereby enhancing the denoising performance in both the simulation
and experimental domains. Experimental results demonstrate that the proposed PDA-Net successfully enhances the quality of STM images, offering promising applications to enhance scientific discovery and accelerate experimental quantum material research.

\end{abstract}

\def\abstractname{Note to Practitioners}
\begin{abstract}
This study is motivated by the challenges of denoising images in scientific fields where acquiring clean ground truth data is infeasible. In material characterization (e.g., STM imaging), evaluating denoising performance is difficult because real experimental images inherently contain noise, and their clean counterparts are unavailable. Fortunately, physics-based simulations can generate an unlimited number of clean STM images, offering a valuable resource for guiding the denoising process.
This paper presents PDA-Net, a physics-augmented deep learning model designed to enhance the denoising of real-world STM images by leveraging simulation data. By integrating adversarial domain adaptation and cycle-consistency modules, PDA-Net effectively transfers knowledge from simulated to experimental domains, enabling improved denoising performance without requiring clean experimental data.
	
\end{abstract}


\section{Introduction}
    Digital imaging techniques have become indispensable across a range of fields, from medical diagnostics and remote sensing to everyday photography. For example, in the medical field, Magnetic Resonance Imaging (MRI) stands as a vital non-invasive tool, providing detailed images of soft tissues, organs, and other internal structures \cite{frisoni2010clinical}. The quality of these images is crucial for accurate disease diagnosis and treatment planning \cite{rutt1996impact,xie2024hierarchical}. However, the presence of noise, especially in low-field MRI or fast imaging protocols, can significantly compromise image clarity, potentially affecting clinical outcomes. Advanced imaging techniques also play a critical role in scientific research. Scanning Tunneling Microscopy (STM), for example, is a powerful method for examining the local electronic structure of quantum materials at atomic resolution \cite{binnig1987scanning}. By leveraging quantitative spectroscopy based on tunneling current, STM enables the visualization of both the geometric and electronic properties of a material’s surface. However, STM is highly sensitive to the distance between the tip and sample, making it susceptible to noise that can obscure crucial details and hinder precise analysis \cite{hofer2003theories}. %Image denoising has become one of the fundamental tasks in image processing, aiming to remove noise from images to restore their original quality.
    To effectively address this challenge, effective image-denoising models are needed to enhance the image quality for better visualization of materials at the atomic or molecular scale.

    %The presence of noise not only affects the visual appearance of the image but also poses challenges for tasks such as key feature extraction and accurate quantitative analysis. 
    Image denoising is a critical preprocessing step in image-based analytics, aimed at removing noise while preserving important image details, such as edges and textures. Recently, deep learning, particularly convolutional neural networks (CNNs) \cite{lecun1998gradient}, has revolutionized the field of image denoising by leveraging large datasets and designing sophisticated architectures. Despite these advancements, image denoising remains a challenging problem. Many denoising models require a substantial amount of paired training data, i.e., clear and blurred pairs, to learn a trustworthy model. Collecting such paired data in real-world settings can be highly challenging. To cope with this limitation, unsupervised image-denoising approaches via Generative Adversarial Networks (GANs) have been increasingly developed such as CycleGAN \cite{zhu2017unpaired}, DualGAN \cite{yi2017dualgan}, Pix2pix \cite{isola2017image}, and UNIT \cite{liu2017unsupervised}. These methods leverage unpaired data to learn mappings between different image domains, achieving unsupervised image denoising. Nevertheless, traditional unsupervised denoising models assume the availability of both noisy and clear images for training. It is important to note that, in many real-world scenarios such as STM, the acquisition of noise-free images is impractical due to the extreme sensitivity to environmental vibrations, thermal drift, electronic noise, and inherent equipment limitations \cite{rerkkumsup2004highly}. This lack of clear images presents a significant challenge for existing models in denoising STM images. 
    
    %Moreover, existing image-denoising models are purely data-driven and are not capable of exploiting the underlying physics-based knowledge to enhance the denoising process.

    %Image blurriness is a major factor that deteriorates image quality and negatively impacts information extraction and pattern recognition for various computer vision applications, such as facial recognition \cite{adjabi2020past, boutros2023synthetic, lu2021human}, object detection \cite{diwan2023object, jiang2022review, ren2015faster}, image style translation \cite{deng2022stytr2, gatys2016image} and 3D reconstruction \cite{tatarchenko2019single}. Image denoising has become one of the fundamental tasks in image processing, aiming to remove noise from images to restore their original quality. 
    
    In this work, we propose a novel physics-augmented adversarial domain adaptive model (PDA-Net) to achieve effective denoising of experimental STM images despite the absence of real-world clear images. Specifically, our contributions are summarized as follows:
   
   (1) We leverage the physical laws governing electron scattering within materials to generate a large volume of unpaired simulated clear and blurry STM images.  The simulated image pairs provide essential clean and noisy feature information as a foundation to construct the denoising model for real-world experimental image denoising.
   
   (2) The novel design of our PDA-Net features with the combination of a cycle-consistency and a domain adversarial module. The cycle-consistency module is proposed to address the challenge of lacking paired training data and ensure the preservation of original image patterns in the denoised outputs. The domain adversarial module is proposed to facilitate simulation-to-real domain adaptation, enabling effective denoising of experimental images using knowledge extracted from simulated images.

   (3) To fully exploit the high similarity between simulated and real experimental images, we propose to implement feature alignment and weight sharing between the two denoising networks for simulated and experimental images, thereby enhancing the denoising performance in both the simulation and experimental domains. 
   
   This effective STM denoising method holds great significance because it will enhance the clarity and accuracy of the atomic-scale structures observed, which is crucial for advancing scientific understanding of material properties, atomic arrangements, and surface phenomena, facilitating new physics discovery in nanotechnology and materials science. The remainder of this paper is organized as follows: Section 2 presents the literature review of image denoising. Section 3 introduces the proposed PDA-Net framework. Section 4 shows the numerical experiments of the proposed PDA-Net method to denoise the simulated blurry images and lab-collected STM images. Section 5 concludes the present investigation.

\section{Literature Review}

    \subsection{Image denoising models}
   Traditional statistical models for image denoising include spatial domain methods and transform domain methods \cite{fan2019brief}. Spatial domain methods operate directly on the pixel values of the image to reduce noise while preserving important features \cite{kumar2013image, shrestha2014image}, %These methods can be further categorized into filtering methods and regularization methods. Filtering methods, such as Gaussian  \cite{kumar2013image},  median \cite{shrestha2014image}, and bilateral filtering \cite{zhang2008multiresolution}, work by modifying pixel values based on the values of neighboring pixels, effectively smoothing the image and reducing noise. 
    %On the other hand, regularization methods \cite{raj2012denoising, jidesh2019estimation}%, like total variation denoising \cite{raj2012denoising} and non-local regularization \cite{jidesh2019estimation}, 
    %formulate denoising as an optimization problem, achieve noise reduction through prior information and smoothness constraints. 
    which have been widely used due to their simplicity and high efficiency in various noise conditions. However, it has been recognized that spatial domain methods have disadvantages in potentially blurring fine details and edges along with noise, leading to a loss of important image information. Transform domain methods were then introduced to better separate noises from signals, leveraging the frequency domain representation of images \cite{ergen2012signal, tian2023multi, you2023research}. These methods suppress certain frequency components of the raw image that may represent noise. Denoising is then achieved by reconstructing the image from the modified frequency domain using the inverse transform, thereby preserving structural details while reducing noise. However, these traditional denoising methods heavily rely on pre-defined mathematical models and assumptions about noise and signal, lacking the flexibility to adapt to varying noise patterns and image structures. 
    
    To meet the growing demand for high-quality images, researchers have engaged deep learning as a powerful alternative. In particular, CNNs have emerged as the most commonly used network architecture to learn complex patterns directly from data \cite{xie2024automated,wang2023hierarchical}, enabling blind image denoising, where no prior knowledge of the blurring function is known \cite{zhang2022deep, min2018blind,huang2020joint, shen2018deep}. Based on CNN backbone, various enhancements, such as perceptual loss \cite{johnson2016perceptual,lu2019uid, zhang2020deblurring}, attention mechanism \cite{vaswani2017attention, zhang2020joint}, multi-scale learning \cite{nah2017deep, kim2022mssnet, dong2023multi}, and vision transformer \cite{fan2022sunet,yi2022masked, tsai2022stripformer}, have been proposed and validated to be effective in improving the denoising performance. Yet, most denoising models require a substantial amount of paired training data (i.e., clear and blurred pair), which are challenging to acquire in real-world scenarios, to learn a trustworthy model. %{\color{red}To suppress the intensive demand for real-world data, many blur generation methods have been proposed, including deep learning-based blur generation \cite{zhang2020deblurring, kupyn2018deblurgan, long2024face, jing2022semantically}, blur kernel \cite{carbajal2023blind,ding2010analysis, chen2020gaussian}. The blur-generation methods aim to create synthetic blurred images from sharp images, facilitating the creation of large datasets required for training. Nonetheless, they are not capable of modeling all types of blur variations encountered in real-world scenarios, which affects the performance of CNN-based methods on real image denoising. (\textbf{Do we really need this part?})}
    
    Recently, GAN-based models \cite{goodfellow2014generative} have emerged as a powerful approach for image denoising, offering significant improvements over traditional deep learning methods. GANs consist of two neural networks, a generator and a discriminator, that are trained simultaneously through adversarial learning. The generator tries to create realistic data and the discriminator attempts to distinguish between real and generated data. %Due to their powerful capability in computer vision \cite{creswell2018generative}, GANs have been utilized to conduct denoising tasks in both supervised and unsupervised manners. In the supervised training, both clear and the corresponding noisy images shall be presented to model training. To address the low feasibility of obtaining paired noisy images, noisy images are either created by manually adding appropriate blur kernels \cite{kupyn2018deblurgan, kupyn2019deblurgan}, or an additional noise generation network \cite{zhang2020deblurring, chen2018image}. The first approach requires extra preprocessing steps, while the second approach can significantly increase model complexity and computational cost.   
     The unique structure of GANs, where the discriminator oversees the images created by the generator, ensures that the generated images possess the expected attributes, i.e., denoised features in this study. This adversarial architecture makes GANs highly suitable for unsupervised image denoising, as they can learn from unpaired data and produce high-quality denoised images without requiring extensive noisy-clear paired datasets. 
     %In an unsupervised setting, GANs leverage the adversarial training framework to learn from unpaired or unlabeled data, significantly easing the data collection and/or synthesis process and enhancing applicability to real-world scenarios. 
     
     A notable example is the CycleGAN framework \cite{zhu2017unpaired}, which employs cycle-consistency loss to ensure that the mapping between noisy and clean image domains is robust, even without paired datasets. Multiple effective denoising approaches have been built upon CycleGAN, demonstrating its effectiveness in producing high-quality denoised images.
    % Techniques such as CycleGAN \cite{zhu2017unpaired} and Noise2Noise \cite{lehtinen2018noise2noise} have been developed to perform image denoising without the need for paired noisy and clean images. 
    For example, Lu et al. \cite{lu2019uid} proposed a disentangled framework to split the content and the blur features of blurred images. They engaged the cycle-consistency model to match the content structures of restored images with the original ones. Zhao et al. \cite{zhao2022fcl} proposed a lightweight unsupervised image deblurring framework with a parameter-free domain contrastive unit to realize more efficient training and inference while preserving the denoising quality.  Kwon \cite{kwon2021cycle} et al. engaged one invertible generator to fulfill the cycle consistency condition instead of two generators, hence increasing the training efficiency. However, unsupervised denoising models in the literature are based on the sense that, although clear and noisy images are not paired, clear images are widely accessible. They did not consider the scenarios where real-world clear images are not available. Even though some work \cite{zhang2024decogan} employs iterative reconstruction algorithms to create clear image benchmarks, this method imposes high computational costs and has a unique limitation of requiring a properly designed prior that accurately captures the characteristics of input images \cite{park2019unpaired}.

      \subsection{Domain Adaptation}
    %In the absence of real-world clear images in the STM domain, we are fortunate to leverage physics knowledge to produce clear impressions of STM images from a simulated perspective. However, due to the high cost of collecting a sufficient number of real-world STM images, traditional unsupervised models may struggle to perform well because of the lack of input variety. To address this, we create a blurred dataset $I_{BS}$ from clear simulated images that resemble the blur patterns observed in real-world data. Our goal is to transfer the knowledge gained from deblurring $I_BS$ to effectively deblur $I_E$, the lab-collected STM images.
    
Domain adaptation \cite{ganin2015unsupervised}, a transfer learning technique, has been increasingly investigated for addressing domain shift and bias, particularly in scenarios with limited or no labeled data. This method enables the distinguish of domain-specific characteristics and the generation of complex samples across diverse domains. For example, Liu et al. \cite{liu2016coupled} proposed coupled GANs to learn a joint distribution of multi-domain images,  facilitating the generation of samples across different domains.
Notably, in the work by Tzeng et al. \cite{tzeng2017adversarial}, the authors proposed the adversarial discriminative domain adaptation model, which utilizes an untied generative mapping to produce output and discriminators to align representations from the source and target domains through adversarial training. Following these pioneering works, many subsequent domain adaptation methods have been proposed with enhanced performance and broader applicability. For instance, the CyCADA model by Hoffman et al. \cite{hoffman2018cycada} introduced a cycle-consistent adversarial approach to adapt representations at both pixel and feature levels, significantly improving the transferability across domains. Lee et al. \cite{lee2021dranet} proposed a content-adaptive domain transfer module, which disentangles image representations into content and style, preserving content structure while effectively transferring style, thereby achieving high-quality domain transfer images.

%Domain adaptation,  allows us to generalize features across different domains, enhancing the model's robustness, which aligns perfectly with our goal of denoising images in the real-world experimental domain based on knowledge learnt from the simulation domain. 

Due to the high quality of feature learning across domains, domain adaptation is also proving to be highly effective in the field of image denoising. For example, Lin et al. \cite{lin2019real} used domain adaptation to map a source noisy image to a target noisy image, both sharing the same ground truth noise-free image, thereby expanding the training set. Li et al. \cite{li2023low} pretrained a modularized adaptive processing neural network, which itself serves as a denoising model, to preprocess training images using paired datasets from multiple domains, thereby enforcing domain adaptation and generating content features with combined knowledge from different domains. %Deng et al. propose an unsupervised domain adaptive image denoising method, which is grounded in the observation that images from similar samples share common content characteristics. 
Deng et al. \cite{deng2024unsupervised} designed a disentanglement-reconstruction network to interchange the noisy pattern and image content between images from two domains. The new images then pass through a denoising generator and are encouraged to match the original paired clear images. However, most existing domain adaptive denoising models either engage the domain adaptation during the preprocessing phase, which still requires clear images to prepare the training inputs, or paired noisy-clean images are demanded to constrain the domain adaptive denoising generator. To the best of our knowledge, we are the first to create an unsupervised domain adaptive denoising model that addresses the scenario where no clear image is available for the target denoising domain.

%Li et al. \cite{li2023low} proposed a GAN with noise encoder transfer learning to generate a paired dataset with different noise styles to address the noise adaptation problem across multiple noise domains. 

%LIN: still know the clear benchmark
%require pretrain model, mix datasets as domain adaptation. with synthesis data set. still need clean labels
%require corresponding clean label

% Zhang et al. \cite{zhang2020joint} proposed propose a supervised network for joint image deblurring and super-resolution. Their model integrates deblurring with channel attention, a super-resolution module, and a dual module to connect the deblurred image with the higher-resolution images.

   	 % It is worth noting that, conventional unsupervised denoising models, which use simulated clear images and lab-collected STM images as training input, may struggle to generate good denoising results due to insufficient variabiliity and image counts in the lab-collected image set.

%%%%%%%%%%%%%%%%%%%%%%%%%%%%%%%%%%%%%%%%%%%%%%%%%%%%%%%%%%%%%%%%%%%%%%%%%%%%%%%%%%%%

\section{Research Methodology}

\begin{figure*}
	\centering
	\includegraphics[width=5.8in]{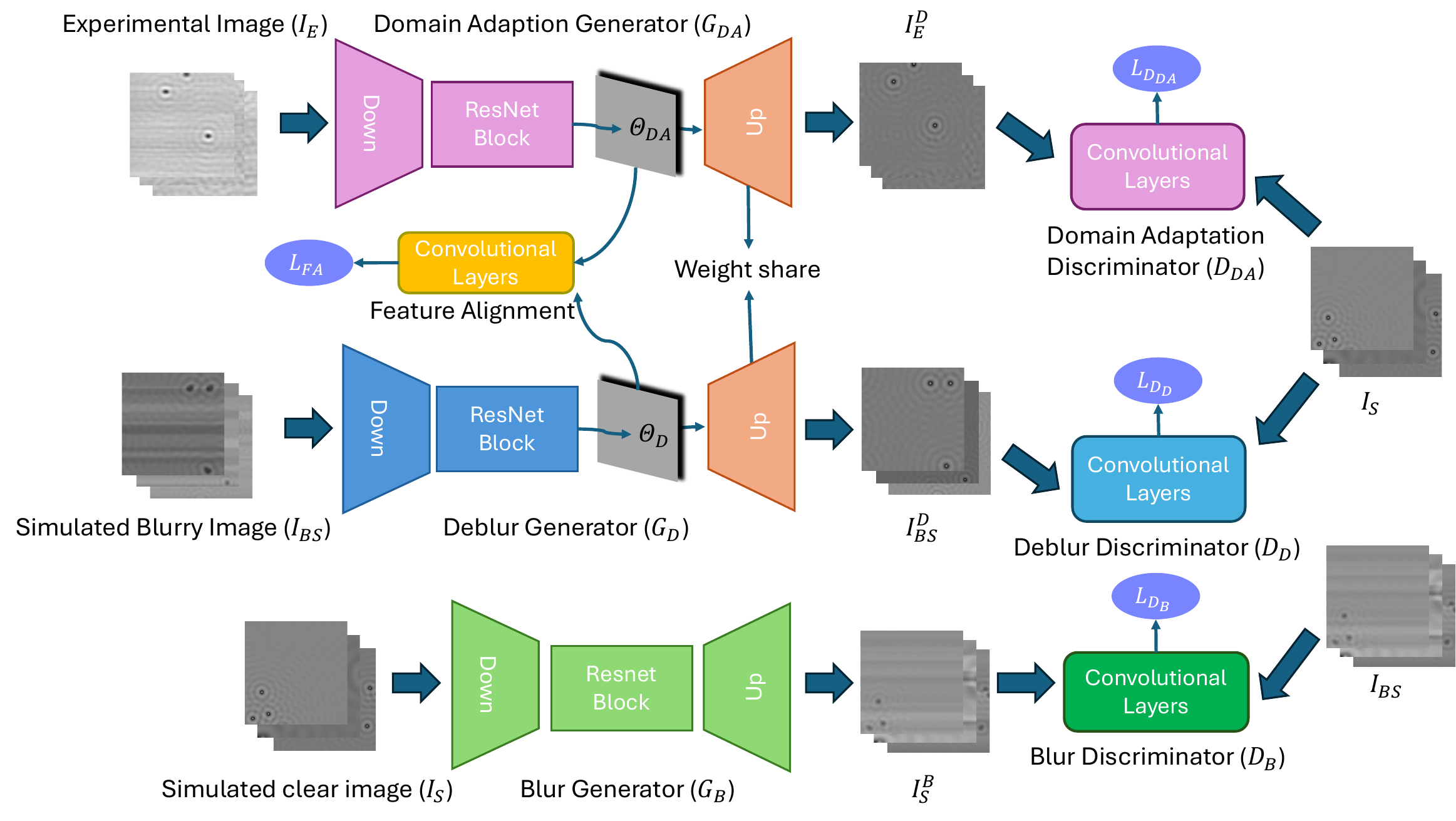}
	\caption{Overview of the PDA-Net model. The PDA-Net model engages three generators. $G_D$ aims to denoise the simulated blurry image $I_{BS}$.  $G_B$, on the contrary, is set to create blurry images from $I_{S}$. $G_{DA}$ is a domain adaptive generator that is specialized to denoise the real-world images $I_E$. The purpose of auxiliary blur generator and discriminator, i.e., $G_B$ and $D_B$ is to formulate a cycle-consistency regularization (see Fig. \ref{Fig:cyc}) and further assist the learning of $G_D$. $D_D$ and $D_{DA}$ are set to discriminate the deblurred images $I_{BS}^D$ and $I_E^D$ with the clear images.}
	\label{Fig:framework}
\end{figure*}

 We propose a PDA-Net for imaging denoising in an unsupervised manner by leveraging physics-based simulation and novel design of network architectures. %Our PDA-Net model aims to fuse the physics-based knowledge into deep learning for denoising experimental images (e.g., STM images) via an innovative adversarial domain adaptation technique. 
 Fig. \ref{Fig:framework} shows the overall flowchart of our PDA-Net model. Here, we use $I_S$ to denote the simulated clean images generated from the physics-based model, $I_{BS}$ to denote the simulated blurry images, and $I_E$ to denote the experimental images. Our goal is to learn a simulation-denoising model  $G_D$ that can generate high-quality deblurred simulated images $I_{BS}^D$ from $I_{BS}$, i.e., $I_{BS}^D = G_D(I_{BS})$, which will be further leveraged to construct an experimental-denoising model $G_{DA}$ that generates high-quality deblurred experimental images $I_E^{DA}$ from $I_E$, i.e., $I_E^{DA} = G_{DA}(I_{E})$. Each component in the flowchart (i.e., Fig. \ref{Fig:framework}) is described in detail in the following subsections.

%The image denoising task is conducted in an unsupervised manner. We are provided with simulated images $I_S$, simulated blurry images $I_{BS}$, and the experiemental images $I_E$. The goal is to learn a model  $G_D$ that can generate deblurred simulated images $I_{BS}^D$ from $I_{BS}$, and a model $G_{DA}$ that generates deblurred experimental images $I_E^{DA}$ from $I_E$, i.e., $I_{BS}^D = G_D(I_{BS})$, and $I_E^{DA} = G_{DA}(I_{E})$. 

\subsection{Adversarial Denoising Model for Physics-based Simulation Images}
\label{section:adversarial}

The simulation denoising model, characterized by a denoising generator $G_D$, aims to sharpen the simulated blurry images $I_{BS}$. To facilitate unsupervised learning and ensure the generated images from $G_D$ look similar to the simulated clear images $I_S$, we introduce an adversarial discriminator, $D_D$, hoping the produced deblurred sample $I_{BS}^D$ by $G_D$ can fool $D_D$. Specifically, $D_D$ aims to differentiate the generated deblurred images $I_{BS}^D$ from the simulated clear images $I_S$, forcing the generator $G_D$ to increasingly generate accurate and realistic deblurred images. This goal is achieved by optimizing the following adversarial objective function:
%We begin by learning a denoising generator $G_D$ that can sharpen the blurred simulated images $I_{BS}$. In order to make the generated images look closer to the clear simulated images $I_S$, we introduce an adversarial discriminator, $D_D$, hoping the produced deblurred sample $I_{BS}^D$ by $G_D$ can fool $D_D$.  Conversely, the adversarial discriminator $D_D$ tries to identify the generated deblurred images $I_{BS}^D$ from the simulated images $I_S$. Thus, the adversarial loss can be formulated as:
\begin{align}
	\label{Eq.DD}
\min_{G_D}\max_{D_D}\mathcal{L}_{D_D} = &\mathbb{E}_{I_{S}\sim p(I_{S})}[\log D_D(I_{S})] \\ \nonumber
&+ \mathbb{E}_{I_{BS}\sim p(I_{BS})}[\log (1-D_D(G_D(I_{BS})))]
\end{align}
In this adversarial game, the generator $G_D$ and discriminator $D_D$ are trained such that $G_D$ seeks to minimize this adversarial loss, while $D_D$ tries to maximize it. This min-max training enables $G_D$ to produce deblurred images $I_{BS}^D$ that are increasingly indistinguishable from the simulated clear image $I_S$, while simultaneously enhancing $D_D$'s ability to discern $I_S$ from the generated ones. It is worth noting that the neural network is trained in an unsupervised manner, meaning there is no direct one-to-one correspondence between the clear image $I_S$ and its blurred counterpart $I_{BS}$ in the training set. 

%In denoising task, there certainly contains significant similarity between the target domain and source domain, ...

    \subsection{Cycle Consistency-enhanced Denoising of Physics-based Simulation Images}
    
    Because no pairwise supervision is provided, the sole unsupervised adversarial denoising model often suffers from instability issue, generating non-ideal deblurred results. This is because, while the adversarial loss in Eq. (\ref{Eq.DD}) encourages the generated deblurred sample $I^D_{BS}$ to resemble the distribution of the clear image $I_S$ simulated by the physics-based model, there is no assurance that $I^D_{BS}$ will maintain the structure or pattern of the original input $I_{BS}$ of the denoising generator $G_D$ \cite{hoffman2018cycada}. To cope with the limitation and ensure the generated deblurred images preserve the patterns of the original inputs, a cycle consistency constraint  \cite{zhu2017unpaired, li2021residual,gu2021adain} is imposed on the generating process. The cycle-consistency constraint is achieved through the following two mechanisms: 
    
    \subsubsection{Adversarial training in the blurred simulated image domain}
     The goal here is to learn a blurring generator $G_B$ that can create blurred images from the simulated clear images $I_S$, i.e., $I^B_S = G_B(I_S)$. Similar to the adversarial model described in \ref{section:adversarial}, a discriminator $D_B$ is deployed to classify the blurry image $I_{BS}$ provided by the physics-based simulation from generated ones $I^B_S$. The adversarial objective function can be written as:
     \begin{align}
         \min_{G_B}\max_{D_B}\mathcal{L}_{D_B} = &\mathbb{E}_{I_{BS}\sim p(I_{BS})}[\log D_B(I_{BS})] \nonumber\\ 
    	&+ \mathbb{E}_{I_{S}\sim p(I_{S})}[\log (1-D_B(G_B(I_{S})))] 
     \label{Eq:DB}
     \end{align}
    Optimizing Eq. (\ref{Eq:DB}) enables $G_B$ to generate blurred images $I_{S}^B$ that are increasingly indistinguishable from the blurred image $I_{BS}$ simulated by the physics-based model, while simultaneously refining $D_B$'s ability to discern $I_{BS}$ from $I_{S}^B$.

    \subsubsection{Cycle-Consistency Loss}
 
    To further enforce the cycle consistency constraint, we need to ensure that the deblurred image $I_{BS}^D$ generated by denoising generator $G_D$ can be reblurred to replicate the original blurred samples, and meanwhile the blurred image $I^B_S$ generated by the blurring generator $G_B$ can be deburred and converted back to clear images. Hence, we can define a forward translation from generated deblurred images to blurred images as:
    $$\hat{I}_{BS}=G_B(I_{BS}^D)$$
     where we expect $\hat{I}_{BS}\approx I_{BS}$. A backward translation from generated blurred images to deblurred images is defined as:
     $$\hat{I}_{S}=G_D(I^B_S)$$
     where the cycle should bring the generated blurred images $I_S^B$ back to their clear version, i.e.,  $\hat{I}_{S}\approx I_{S}$. The cycle consistency loss is defined as the L1 regularization on the reconstruction errors of both translations:
     \begin{align}
     	& \mathcal{L}_{cyc} = \mathcal{L}_{cyc}^f + \mathcal{L}_{cyc}^b \\
     	\text{where} \quad  &  \mathcal{L}_{cyc}^f= \mathbb{E}_{I_{BS}\sim p(I_{BS})}[\|I_{BS}-\hat{I}_{BS}\|_1]  \nonumber\\
     	&\mathcal{L}_{cyc}^b =  \mathbb{E}_{I_S\sim p(I_S)}[\|I_S-\hat{I}_S\|_1] \nonumber
     \end{align}
     
    Minimizing the cycle-consistency loss will further narrow down the distribution of the generated images and preserve the patterns of original images, leading to enhanced performance in imaging denoising in the simulation domain.
    
     \begin{figure}[h]
   	\centering
   	\includegraphics[width=2.5in]{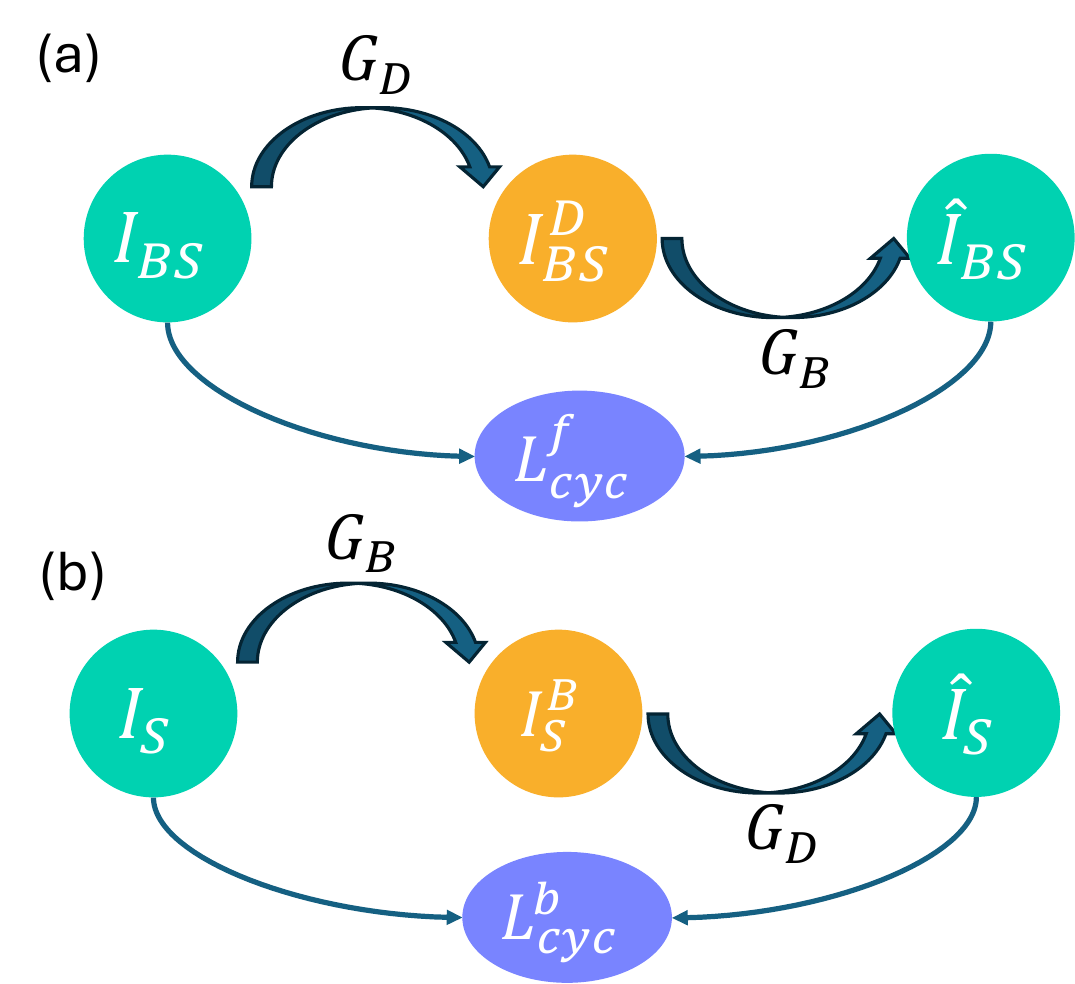}
   	\caption{Given the intuition that translating images from one domain to another and then reversing the process should restore the original images, the cycle is expected to return the generated images to their original state, we design (a) a forward consistency translate with loss $\mathcal{L}_{cyc}^f$ and (b) a backward consistency translate with loss $\mathcal{L}_{cyc}^b$. }
   	\label{Fig:cyc}
   \end{figure}

    \subsection{Simulation-to-Real Domain Adaption for Experimental Image Denoising }

    The simulation-denoising model $G_D$ is expected to provide good denoising performance for simulated images, leveraging the advantages of adversarial objectives and the cycle-consistency constraint. However, directly applying $G_D$, trained on simulated images, to denoise real experimental images, denoted as $I_E$, is inappropriate and may generate suboptimal results due to the domain shift issue. Training an experimental-denoising model from scratch using real-world experimental images is generally infeasible because it is extremely difficult or expensive to collect noise-free experimental images. To cope with this challenge, this subsection introduces a Simulation-to-Real Domain Adaption method to adapt the simulation-denoising model $G_D$ such that it performs well on denoising real-world experimental data without any additional supervision from noise-free experimental images.
    
    %We have so far described a unsupervised denoising model, which combines adversarial objectives and cycle consistency. The denoising generator $G_D$ is trained on unpaired simulated clear images $I_S$ and simulated blurry images $I_{BS}$. Our primary objective, however, is to denoise real experimental images obtained from scanning electron microscopy (STM), i.e., $I_E$. Unlike the simulated images, there are no clear benchmarks for $I_E$, making the task more challenging. There exsist significant discrepancies in distribution and noise characteristics between the simulated blurry images used for training and the real-world experimental images. For example, the quality of experimental images obtained from STM is strongly impacted by STM parameters such as accelerating voltage, stigmation and focus, working distance, and other random artifacts, which certainly fall outside the capabilities of the blurry simulator. As such, applying $G_D$ directly to $I_E$ could lead to suboptimal results. This model limitation highlights the necessity for domain adaptation, prompting us to develop a specialized denoising generator tailored for deblurring the experimental images $I_E$. %The proposed PDA-Net model will also take advantage of simulated clear and blurry images, i.e., $I_S$ and $I_{BS}$, to learn the underlying transformations necessary for effective deblurring, eventually enhancing generalization of real-world STM images.
    
    For implementing the domain adaptation, we first create a base generator $G_{DA}$, which inherits the same neural network structure with $G_D$. However, the neural network parameters in $G_{DA}$ used for deblurring real-world images $I_E$ will differ from those in $G_D$ due to the unavoidable domain shift between the simulation and real experimental domains. Thus, it is necessary to retrain $G_{DA}$ and fine-tune the neural network parameters specifically for the experimental image domain to ensure effective performance. Here, we propose to achieve domain adaptation and optimize $G_{DA}$ in an adversarial manner by incorporating a domain discriminator $D_{DA}$. Specifically, $D_{DA}$ evaluates the deblurred experimental images, denoted as $I_E^D = G_{DA}(I_E)$, and compares them against the clear simulated images $I_S$  by continuously challenging $G_{DA}$ to produce deblurred outputs whose sharpness degree closely resembles that in $I_S$. The adversarial domain-adaptation objective function is defined as:
    \begin{align}
        \label{Eq:DA}
    	\min_{G_{DA}}\max_{D_{DA}}\mathcal{L}_{D_{DA}} &= \mathbb{E}_{I_S \sim p(I_S)}[\log D_{DA}(I_S)] \nonumber \\
    	&\quad + \mathbb{E}_{I_E \sim p(I_E)}[\log (1 - D_{DA}(G_{DA}(I_E)))]
    \end{align}  
    
    The proposed domain adaption technique aims to eliminate the domain discrepancy between the simulated and real experimental images such that the denoising process learned from simulated images can be readily applicable to denoising the real-world experimental images. However, using a separate adversarial generator $G_{DA}$ for our target domain, i.e., experimental imaging, without any connection to the simulation domain, may require a larger amount of data from both domains and a longer training time to achieve good performance, as it must independently learn all necessary features from the ground up. Additionally,  isolating $G_{DA}$ from $G_D$ gives up the opportunity to utilize valuable, domain-invariant features learned from the simulation domain, potentially reducing the effectiveness of $G_{DA}$ in denoising images from the real experimental domain. 
    
   Given that both the simulated images, derived using physics laws, and the experimental images depict the same physical subject (e.g., Cu atoms in STM experiments), there should inherently be substantial pattern similarities between the images from the two domains. Such similarities need to be effectively explored and incorporated into the generators $G_D$ and $G_{DA}$ to improve the denoising performance of the model. Hence, we propose to leverage the weight sharing and feature alignment mechanisms to enable the two generators to incorporate the similarity information from the simulation and real experimental domains. 
   
   %By implementing partial weight sharing, we facilitate the direct transfer of learned features from the simulated to the experimental domain. Similarly, feature alignment helps ensure that the critical features from both domains are processed in comparable ways, enhancing the model's ability to generalize across different sourced image data. This approach effectively reduces the redundancy in learning similar features separately for each domain, optimizing both the performance and data utilization of the model. % These strategies not only can expedite the learning process but also ensures consistency in feature representation across two domains. %, reducing the overall complexity and resource requirements of the model. %These strategies are key in enhancing the model's generalizability and efficiency, making the domain adaptation process more seamless and effective.
   
   \subsubsection{Weight sharing}
   As depicted in Fig. \ref{Fig:framework}, $G_D$ and $G_{DA}$ share the same network structure, i.e., an encoder-decoder architecture with a ResNet block at the bottleneck. %This structure consists of an encoder for downsampling, a central ResNet Block for enhanced feature extraction, and a decoder for upsampling to reconstruct the original resolution. 
   We assume that the Down and ResNet Block components within $G_D$ and $G_{DA}$ are trained independently, enabling each to learn domain-specific attributes for effectively denoising the respective input images, i.e., $I_{BS}$ and $I_E$. The weight sharing is then implemented in the Upsampling blocks in  $G_D$ and $G_{DA}$, tasking with further purifying the common noisy patterns that appear in both domains and constructing feature maps to the original image resolution. This strategy facilitates the transfer of learned representations from the simulation to the experimental domain, ensuring that both generators benefit from shared knowledge and reduce redundancy in learning. Based on the premise that images from two distinct domains exhibit similar high-level attributes, as they both characterize the same subject (i.e., copper single crystal in our later case study), the upsampling blocks in  $G_D$ and $G_{DA}$ are configured to share the same weights:
   \begin{equation}
   	\theta_{DA}^{up} = \theta_{D}^{up}
   \end{equation}
   where $\theta_{DA}^{up} $ and $\theta_{D}^{up}$ are the neural network parameters for upsampling blocks (orange) in $G_D$ and $G_{DA}$, respectively.

   \subsubsection{Feature Alignment}
   We further engage feature alignment to synchronize the intermediate features that are capable of characterizing the overall input pattern. The individually trained Down and ResNet blocks in both generators are expected to extract blurring information from their own corresponding domains. 
   This allows each generator to precisely address the unique denoising challenges posed by different noise sources—simulated blurring and real-world noisy experimental images, respectively. Consequently, the intermediate feature maps $\Theta_D = \phi_{G_D}(I_{BS})$ and $\Theta_{DA} = \phi_{G_{DA}}(I_E)$ should exhibit considerable resemblance to each other after the initial denoising step posed by the Down and ResNet Blocks $\phi_{G_{DA}}$ and $ \phi_{G_D}$ from $G_D$ and $G_{DA}$, respectively. As such, we define the following discrimination loss function %({\color{red} Do you need an additional generator here to achieve adversarial training?}) 
   to achieve feature alignment:
   %We assume the Down and ResNet Block components are trained separately for $G_D$ and $G_{DA}$, allowing each to learn specialized attributes for denoising $I_{BS}$ and $I_E$, respectively. As such, the preceding Down and ResNet blocks in $G_D$ and $G_{DA}$ are tailored to extract blurring information from the blurry images of the simulation and experimental domains. This setting allows each generator to learn specialized attributes for denoising the artifacts brought by different noise sources, i.e., simulated bluration and real-world STM blurry artifacts. Consequently, the intermediate feature maps $\Theta_D$ and $\Theta_{DA}$ should exhibit considerable resemblance to each other after the initial denoising step posed by the Down and ResNet Blocks. Thus, the loss function for feature alignment can be written as:   
   %This is because the preceding Down and ResNet Blocks in $G_D$ and $G_{DA}$ have been tailored to extract blurring information from the blurry images of the simulation and experimental domains. Consequently, the intermediate feature maps $\Theta_D$ and $\Theta_{DA}$ should exhibit considerable reSTMblance to each other after the initial denoising step posed by the Down and ResNet Blocks. Thus, the loss function for feature alignment can be written as:
\begin{align}
	\min_{\phi_{G_{DA}}, \phi_{G_{D}}}\max_{D_{FA}} \mathcal{L}_{FA} &= \mathbb{E}_{I_E \sim p(I_E)} \left[ \log D_{FA}(\phi_{G_{DA}}(I_E)) \right] \nonumber \\
	&\quad + \mathbb{E}_{I_{BS} \sim p(I_{BS})} \left[ \log (1 - D_{FA}(\phi_{G_D}(I_{BS})) \right]
\end{align}
where $D_{FA}$ is a domain classifier. %and $\Theta_D= \phi_{G_{DA}}(I_E)$ and $\Theta_{DA}= \phi_{G_D}(I_{BS})$. $\phi_{G_{DA}}$ and $ \phi_{G_D}$ represent the Down and ResNet blocks for $G_D$ and $G_{DA}$, respectively. 

  By implementing partial weight sharing in $G_D$ and $G_{DA}$, we facilitate the direct transfer of learned features from the simulated domain to the experimental domain. Additionally, feature alignment helps ensure that the critical features from both domains are processed in comparable ways, enhancing the model's ability to generalize across different sourced image data. This approach effectively reduces the redundancy in learning similar features separately for each domain, optimizing both the performance and data utilization of the model.

Putting together, the final objective function is the weighted sum of all losses mentioned above:
\begin{eqnarray}
    &\mathcal{L}_{total} \nonumber \\
    =&\lambda_{D} \mathcal{L}_{D_D}+\lambda_{B} \mathcal{L}_{D_B} + \lambda_{cyc} \mathcal{L}_{cyc} +  \lambda_{DA}  \mathcal{L}_{DA} + \lambda_{FA} \mathcal{L}_{FA}
    \label{Eq: final_obj}
\end{eqnarray}
 where $\lambda_{D}$, $\lambda_{B}$, $\lambda_{cyc}$, $\lambda_{DA} $, and $\lambda_{FA}$ denote the weights characterizing the contribution of the corresponding loss term, which is selected by empirically fine-tuning. Optimizing Eq. (\ref{Eq: final_obj}) will provide both the simulation-denoising and experimental-denoising models:
      $\{G_D^*, G_{DA}^*\}=\arg\min_{\{G_D, G_B, G_{DA}, \phi_{G_D},\phi_{G_{DA}}\}}\max_{\{D_D, D_B, D_{DA}, D_{FA}\}}\mathcal{L}_{total}$, where $\phi_{G_D} \subset G_D$, $\phi_{G_{DA}} \subset G_{DA}$.

  % {\bf \color{red}Please provide an algorithm table to illustrate the training process of the model.}
   %As such, the preceding Down and ResNet blocks in $G_D$ and $G_{DA}$ are tailored to extract blurring information from the blurry images of the simulation and experimental domains. This setting allows each generator to learn specialized attributes for denoising $I_{BS}$ and $I_E$. Consequently, the intermediate $\Theta_D$ and $\Theta_{DA}$ should exhibit considerable reSTMblance to each other after the initial denoising step posed by the Down and ResNet Blocks. Thus, the loss function for feature alignment can be written as:

  \subsection{Training Procedure of the PDA-Net}
  
\begin{algorithm}
	\caption{Training Procedure of the proposed PDA-Net model.}
	\begin{algorithmic}
		\Statex \textbf{Input:} $\{I_{BS}\}_{b=1}^{N_b}$, $\{I_S\}_{b=1}^{N_b}$, $\{I_E\}_{b=1}^{N_b}$, $N_b$
		\Statex \textbf{Output:} Trained denoising models: $G_D^*$ and $G_{DA}^*$ %network parameters $\mathcal{W} = \{\mathcal{W}_{D_D}, \mathcal{W}_{D_B}, \mathcal{W}_{D_{DA}}, \mathcal{W}_{D_{FA}}, \mathcal{W}_{G_D}, \mathcal{W}_{G_B}, \mathcal{W}_{G_{DA}}\}$
		
		\For{$n = 1$ \textnormal{to maximum epoch} $N_e$}
		
		\For {$b=1$ \textnormal{to maximum batch number} $N_b$}
		
		\State \parbox[t]{210pt} {Generate a batch of synthetic images $\{I_E^D\}_b$, $\{I_{BS}^D\}_b$, and $\{I_S^B\}_b$ using generators $G_D$, $G_B$, $G_{DA}$.  }
		\vspace{0.5pt}
		%\State Compute the discriminators' losses: $\mathcal{L}_D = \mathcal{L}_{D_D} + \mathcal{L}_{D_B} + \mathcal{L}_{D_{DA}} + \mathcal{L}_{D_{FA}} $.
		\State \parbox[t]{210pt} {Compute the discriminators' losses: $\mathcal{L}_D = \lambda_{D}\mathcal{L}_{D_D} + \lambda_{B}\mathcal{L}_{D_B} + \lambda_{DA}\mathcal{L}_{D_{DA}} + \lambda_{FA}\mathcal{L}_{FA} $.}
		\vspace{0.9pt}
		\State \parbox[t]{210pt}  {Update the discriminator parameters $ \{\mathcal{W}_{D_D}, \mathcal{W}_{D_B}, \mathcal{W}_{D_{DA}}, \mathcal{W}_{D_{FA}}\}$ by descending the gradient of $-\mathcal{L}_D$. Meanwhile, the network parameters for all the generators are frozen.}
		\vspace{0.9pt}
		\State \parbox[t]{210pt} {Compute the generators' losses and cycle-consistency loss: $\mathcal{L}_{G+C} = \lambda_{D}\mathcal{L}_{G_D} + \lambda_{B}\mathcal{L}_{G_B} + \lambda_{DA}\mathcal{L}_{G_{DA}} + \lambda_{cyc}\mathcal{L}_{cyc} + \lambda_{FA}\mathcal{L}_{FA}$.}
		\vspace{0.9pt}
		\State \parbox[t]{200pt}  {Update the generator parameters $\{\mathcal{W}_{G_D}, \mathcal{W}_{G_B}, \mathcal{W}_{G_{DA}}, \mathcal{W}_{\phi_{G_{D}}}, \mathcal{W}_{\phi_{G_{DA}}} \}$ by descending the gradient of $\mathcal{L}_{G+C}$. Note that $\mathcal{W}_{\phi_{G_{D}}} \subset \mathcal{W}_{G_{D}}$, $\mathcal{W}_{\phi_{G_{DA}}} \subset \mathcal{W}_{G_{DA}}$. Meanwhile, the network parameters for all the discriminators are frozen.  
		}     
		\vspace{2.5pt}    
		%\State Update the generator parameters by descending the gradient of $\mathcal{L}_{G+C}$. Meanwhile, the network parameters for all the discriminators are frozen.
		% \State \textbf{end}
		
		\EndFor  
		\If{converge}
		\State \textbf{break} 
		\EndIf
		
		%	      \State \hspace{14pt}\textbf{end}
		
		\EndFor
		
		\State \Return \parbox[t]{210pt} {\raggedright $G_D^*$, $G_{DA}^*$, $\mathcal{W}= \{\mathcal{W}_{D_D}, \mathcal{W}_{D_B}, \mathcal{W}_{D_{DA}}, \mathcal{W}_{D_{FA}}, \mathcal{W}_{G_D}, \mathcal{W}_{G_B}, \mathcal{W}_{G_{DA}}\}$}
	\end{algorithmic}
	\label{alg}
\end{algorithm}

            %Compute the discriminators' losses: $\mathcal{L}_D = \mathcal{L}_{D_D} + \mathcal{L}_{D_B} + \mathcal{L}_{D_{DA}} + \mathcal{L}_{D_{FA}} $.
                
             %Update the discriminator parameters by descending the gradient of $\mathcal{L}_D$.
             %Compute the generators' losses and cycle-consistency loss: $\mathcal{L}_G = \mathcal{L}_{G_D} + \mathcal{L}_{G_B} + \mathcal{L}_{G_A} + \mathcal{L}_{G_{DA}} + \mathcal{L}_{cyc}$

  Algorithm \ref{alg} summarizes the training process for the proposed PDA-Net. For each epoch, and within each training batch, the algorithm generates synthetic denoised and blurry images using a set of generators $G_D$, $G_B$, $G_{DA}$. The algorithm then updates the discriminator parameters by descending the gradient of the discriminator loss $-\mathcal{L}_D$, where $\mathcal{L}_D$ is computed by summing the weighted individual losses of $\mathcal{L}_{D_D}$, $\mathcal{L}_{D_B}$, $\mathcal{L}_{D_{DA}}$, and $\mathcal{L}_{D_{FA}}$. After updating the discriminators, the parameters of generators are updated by descending the gradient of the combined generator and cycle-consistency loss $\mathcal{L}_{G+C}$, which also includes a set of weighted losses ($\mathcal{L}_{G_D}$, $\mathcal{L}_{G_B}$, $\mathcal{L}_{G_{DA}}$, $\mathcal{L}_{cyc}$, and $\mathcal{L}_{FA}$). The generators ${G_D}$ and ${G_{DA}}$ try to produce clear images that are indistinguishable from simulated clear data $I_S$, while $G_B$ aims to yield blurry images that are similar to the simulated blurry ones $I_{BS}$. $\phi_{G_{D}}$ and $\phi_{G_{DA}}$, the Down and Resnet blocks in $G_D$ and $G_{DA}$, strive to align the intermediate feature maps $\Theta_D$ and $\Theta_{DA}$ that are indistinguishable to $D_{FA}$. %Therefore, it aims to fool the discriminators into classifying generated images as simulated. 
  According to the loss functions in Eq. (\ref{Eq.DD}), Eq. (\ref{Eq:DB}), and Eq. (\ref{Eq:DA}), the generator losses are formulated as:
\begin{align}
 \mathcal{L}_{G_D} &= \mathbb{E}_{I_{BS}\sim p(I_{BS})}[\log (1-D_D(G_D(I_{BS})))]   \nonumber\\
 \mathcal{L}_{G_B} &=  \mathbb{E}_{I_{S}\sim p(I_{S})}[\log (1-D_B(G_B(I_{S})))]  \nonumber\\
  \mathcal{L}_{G_{DA}} &=  \mathbb{E}_{I_E \sim p(I_E)}[\log (1 - D_{DA}(G_{DA}(I_E)))]
\end{align} 

This iterative process continues until convergence. The algorithm then returns the trained network parameters $\mathcal{W}$ and the two denoising models: $G_D^*$ and $G_{DA}^*$. During the inference, the testing inputs are passed through $G_D^*$ and $G_{DA}^*$ to produce a deblurred version of simulated blurry images and real-world experimental images.

\section{Experimental Design and Results}

In quantum material research, STM has been a powerful experimental probe for delineating electronic structures at sub-nanometer scales \cite{binnig1987scanning}. However, environmental factors such as mechanical vibration, electrical noise, and thermal fluctuations will inevitably generate noise in the STM data \cite{wang1995stable}, fundamentally limiting its spatial resolution. While the noise reduction of STM data is generally challenging with conventional methods, it provides a meaningful real-world example to test the denoising capability of the proposed PDA-Net architecture.            

\subsection{Physics-based Model to Generate Simulated STM Images}

Our physical system of interest is (111) surface of copper single crystals, i.e., Cu(111), which hosts a Shockley-type surface state described by an effective Hamiltonian operator,
\begin{equation}
    \hat{H}_0 = \frac{{\bm k}^2}{2m_\text{eff}} - \mu.
\end{equation}
Here, ${\bm k}=(k_x, k_y)$ is the crystal momentum of surface electrons. The chemical potential for Cu(111) is $\mu=0.45$ eV, and $m_\text{eff} = 0.38 m_e$ is the effective mass of surface electrons, where $m_e$ denotes the bare electron mass. When an atomically sharp STM tip approaches the surface of Cu(111), it injects an electron into Cu(111) through the quantum tunneling effect. The injected electron will propagate across the surface like a plain wave until it scatters with lattice impurities or defects. The scattered electron wave will superpose with the injected wave, forming an interference pattern and sending a signal back to the STM probe. Moving the STM probe across the sample and performing measurements repeatedly will generate local density of states (LDOS) images that map out the electron density distribution of Cu(111) \cite{crommie1993imaging}. 

We exploit the Green's function method to simulate the above physical process \cite{fiete2003colloquium}. Specifically, we use $\hat{V}({\bf r})$ to denote the potential function that describes $N$ point-like impurities randomly distributed on the surface. The Hamiltonian of Cu(111) with defects will then be updated to $\hat{H}=\hat{H}_0 + \hat{V}({\bf r})$, with the corresponding stationary Schr{\"o}dinger equation as $\hat{H}|\psi_\alpha({\bf r})\rangle = E_\alpha |\psi_\alpha({\bf r})\rangle$. Here, $|\psi_\alpha ({\bf r})\rangle$ is the electron eigenstate with an energy $E_\alpha$, where $\alpha$ is a state index. Hence, the retarded Green's function of the system at a frequency $\omega$ is defined as
\begin{equation}
   {\cal G}({\bf r}',{\bf r},\omega) = \sum_\alpha \frac{\psi_\alpha^*({\bf r})\psi_\alpha({\bf r}')}{\omega - E_\alpha + i \eta},
\end{equation}
where $\eta$ is an infinitesimal positive number. The LDOS function ${\cal A}({\bf r},\omega)$ is then related to the Green's function via 
\begin{equation}
    {\cal A}({\bf r},\omega) = - \frac{1}{\pi} \text{Im} {\cal G}({\bf r}',{\bf r},\omega)
\end{equation}
which is used to generate the simulated STM images.

%Notably, the exact form of ${\cal G}({\bf r}',{\bf r},\omega)$ is often challenging to obtain. Nonetheless, Dyson's equation offers a powerful formalism to evaluate ${\cal G}({\bf r}',{\bf r},\omega)$ in a perturbative manner, with 
%\begin{equation}
%    {\cal G} = {\cal G}_0 + {\cal G}_0 \hat{V} {\cal G}_0 + {\cal G}_0 \hat{V} {\cal G}_0 \hat{V} {\cal G}_0 + ...
%\end{equation}
%The zeroth-order ${\cal G}_0({\bf r}',{\bf r},\omega)$ denotes the impurity-free Green's function whose analytical form is known in the literature. For weak impurities, the above infinite series will rapidly converge, with which we will feasibly achieve accurate simulations of the surface LDOS for Cu(111).}

\subsection{STM Setup to Generate Experimental STM Images}

In the physical experiments, we prepared copper samples with atomically flat (111) surfaces, i.e., Cu(111), inside the ultra-high-vacuum (UHV) chamber using a sputtering-annealing cycles \cite{marks1991uhv}. To acquire real-world experimental images, we used a low-temperature STM that operates at the pressure of $< 10^{-10}$ torr and the base temperature of 4.2 K. The Cu(111) sample was scanned by the STM at a bias voltage of 10 mV and a current set point of 1 nA, which is the parameter setting optimized to achieve high-resolution imaging of the surface features.  50 experimental STM images were acquired to test the performance of our denoising model.

\subsection{Neurual Network Architecture and Optimization}
The generators $G_{DA}$,  $G_D$, and $G_B$ share the same neural network architecture, as shown in Fig. \ref{Fig:framework}. Specifically, the Down block consists of three convolutional layers. The first convolutional layer has a kernel size of $7\times 7$, a stride of 1, a reflection padding of 3. The subsequent two convolutional layers have a kernel size of $3\times 3$, a stride of 2, and a padding of 1. The ResNet block comprises nine residual connection blocks, each containing two convolutional layers with a kernel size of $3\times 3$, a stride of 1, and reflection padding of 1. The Up block reconstructs the feature map back to the original image size. It mirrors the Down block, consisting of two deconvolutional layers with a kernel size of $3\times 3$, a stride of 2, and a padding of 1, followed by a final convolutional layer with a kernel size of $7\times 7$, a stride of 1, a reflection padding of 3. The discriminators $D_D$, $D_B$, and $D_{DA}$ adopt the PatchGAN discriminator architecture without modification \cite{isola2017image}. The weighting hyperparameters that determine the contribution of different loss terms play a crucial role in training PDA-Net; we empirically set $\lambda_D=\lambda_B=\lambda_{cyc}=\lambda_{DA} =1$, while $\lambda_{FA}=0.1$.
The PDA-Net is optimized using the Adam optimizer with a default learning rate of 0.0002 for all network training. The momentum parameter is set to 0.5.  Training is conducted on an NVIDIA A100 GPU with 80 GB memory, ensuring efficient computation and handling of large-scale datasets.

\subsection{Performance evaluation}
During the training process, we prepared 3,600 images each for simulated clear images ($I_S$) and simulated blurry images ($I_{BS}$).The 61 experimental images are split into training and testing sets, with 50 images undergoing augmentation (flipping, rotating, and cropping) to create a training set of 3,600 images ($I_E$). The remaining 11 images are augmented to generate 200 test images. 
%The 50 experimental images ($I_E$) are augmented by applying flipping, rotating, and cropping techniques, resulting in a final dataset with 3,600 images. 
For quantitative evaluation, the test set consists of 200 simulated blurry images ($T_{BS}$) and 200 experimental images ($T_E$) are passed through the trained generators $G_D^*$ and $G_{DA}^*$ for denoising. 

%The denoising performance of our PDA-Net will be benchmarked with a pure cycle-consistence model (i.e., CycleGAN), CycleGAN + Domain Adaptation (i.e., CycleGAN+DA), and CycleGAN+DA + Weight Sharing (i.e., CycleGAN+DA+WS) methods.
We employed three full-reference quality metrics—$MSE$, $PSNR$, and $SSIM$—which compare denoised simulated blurry images in the test set $T_{BS}^D$ to the original clear images. To quantify the quality of denoised experimental images from the test set ($T_E^D$), we use two no-reference metrics—$PIQE$ and $BRISQUE$—to assess the image naturalness. Their definitions are described below: 

\begin{itemize}
	\item  \textit{Mean Squared Error (MSE)}: $MSE$ measures the average squared difference between corresponding pixels of two images, often used to quantify the difference between a reference image and a distorted image. 
	\begin{equation}
		MSE=\frac{1}{m n} \sum_{i=1}^m \sum_{j=1}^n[T_{BS}^D(i,j)-T_S(i, j)]^2
	\end{equation}
	where $m\times n$ is the image size, $T_S$ is the clear simulated image,  and $T_{BS}^D$ is the denoised simulated image generated by our model, i.e., $T_{BS}^D = G_D^*(T_{BS})$, where $T_{BS}$ is simulated blurred image from $T_S$ in test set. 
	
	\item \textit{Peak Signal-to-Noise Ratio (PSNR)}\cite{zhang2015wavelet}: $PSNR$ is defined as a ratio that compares the maximum possible pixel value of an image to the power of noise: 
	\begin{equation}
		PSNR=10 \log _{10}\left(\frac{255^2}{MSE}\right)
	\end{equation}
    A higher PSNR value indicates better image quality, with less distortion relative to the reference image.
    
    \item \textit{Structural Similarity Index (SSIM)}: While MSE and PSNR are widely used for image quality assessment, they have limitations in capturing perceptual differences as they primarily measure pixel-wise errors. As such, we employ SSIM, which offers a more perceptually accurate evaluation by considering luminance, contrast, and structural information \cite{wang2004image}. It is mathematically defined as:
    \begin{equation}
    	SSIM(T_{BS}^D, T_S)=\frac{\left(2 \mu_{D} \mu_{{S}}+C_1\right)\left(2 \sigma_{DS}+C_2\right)}{\left(\mu_D^2+\mu_S^2+C_1\right)\left(\sigma_D^2+\sigma_S^2+C_2\right)}
    \end{equation}
    where and $\mu_D$ and $\mu_S$ are the means of images of $T_{BS}^D$ and $T_S$, $\sigma_D$ and $\sigma_S$ are the corresponding variances, $\sigma_{DS}$ is the covariance, and $C_1$ and $C_2$ are constants to stabilized the division.  SSIM values range from -1 to 1 with higher values indicating better image quality.
    
    \item \textit{$PIQE$ and $BRISQUE$}:  $PIQE$ (Perception-based Image Quality Evaluator) \cite{venkatanath2015blind} and $BRISQUE$ (Blind/Referenceless Image Spatial Quality Evaluator) \cite{mittal2012no} are image quality assessment models without needing a reference image. Both models utilize natural scene statistics to analyze the inherent image properties, capturing deviations from expected natural characteristics. $PIQE$ focuses on detecting perceptually significant distortions based on a block-wise analysis, where it divides the image into smaller blocks and evaluates the quality of each block separately. In contrast, $BRISQUE$ employs a regression model, typically a support vector machine (SVM), to predict quality scores from spatial domain features. %$PIQE$ and $BRISQUE$ are powerful tools for assessing image quality without reference images, with $PIQE$ focusing on local distortions and $BRISQUE$ leveraging machine learning techniques for quality prediction. 
    For both metrics, lower scores indicate better image quality.

\end{itemize} 

   \begin{figure*}
	\centering
	\includegraphics[width=6in]{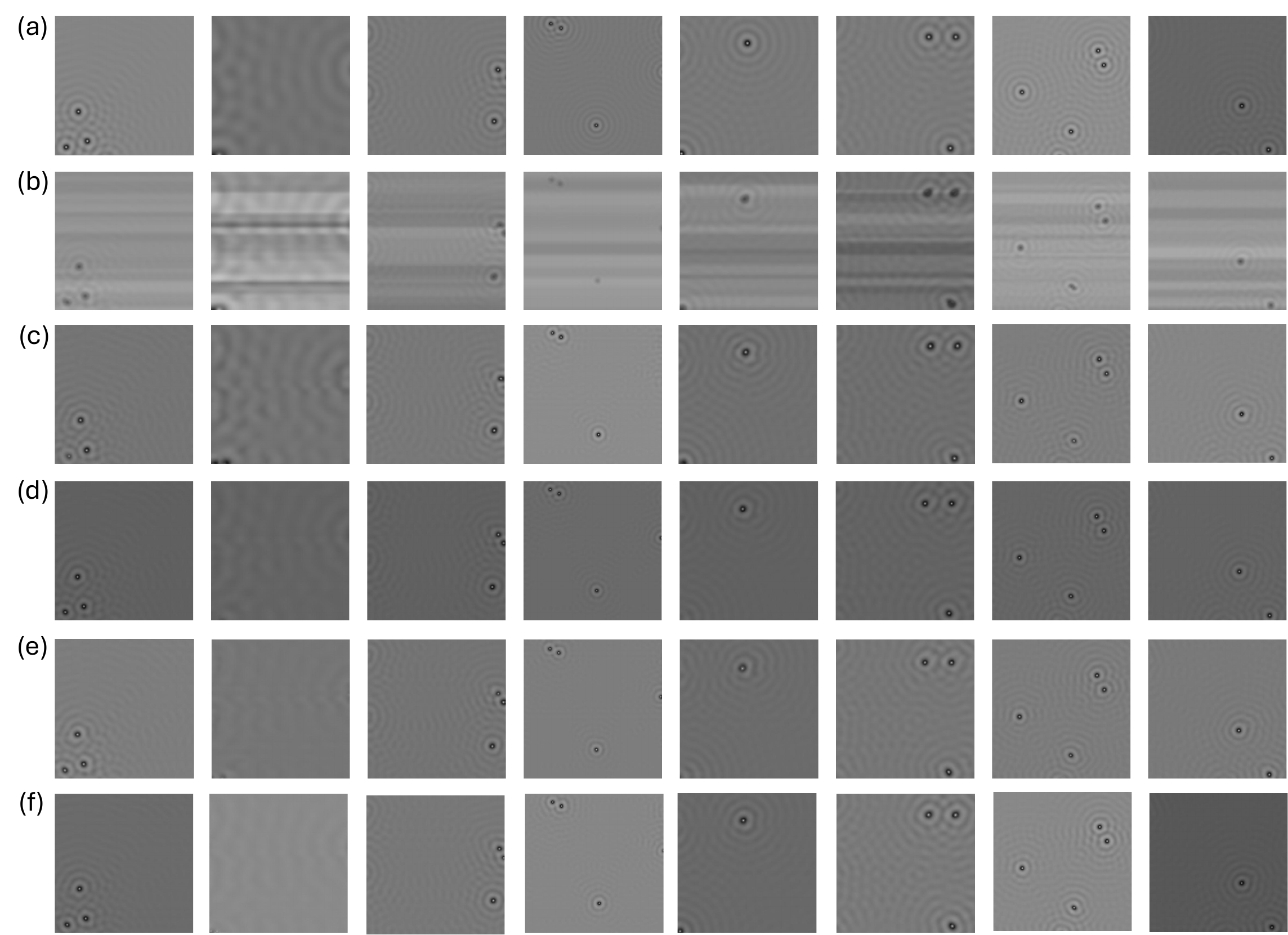}
	\caption{Examples of (a) The simulated clear images ($I_S$); (b) The simulated blurry images ($I_{BS}$); (c) image deblurred from $I_{BS}$ using pure CycleGAN model; (d) images deblurred from $I_{BS}$ using CycleGAN with domain adversarial (DA) module; (e) images deblurred from $I_{BS}$ using CycleGAN and DA with weight sharing; (f) images deblurred from $I_{BS}$ using the proposed PDA-Net model, i.e., $I_{BS}^D$. %{\color{red}Model names should be consistent.}
 }
	\label{Fig:sim_result}
\end{figure*}

 \begin{figure*}
	\centering
	\includegraphics[width=5.2in]{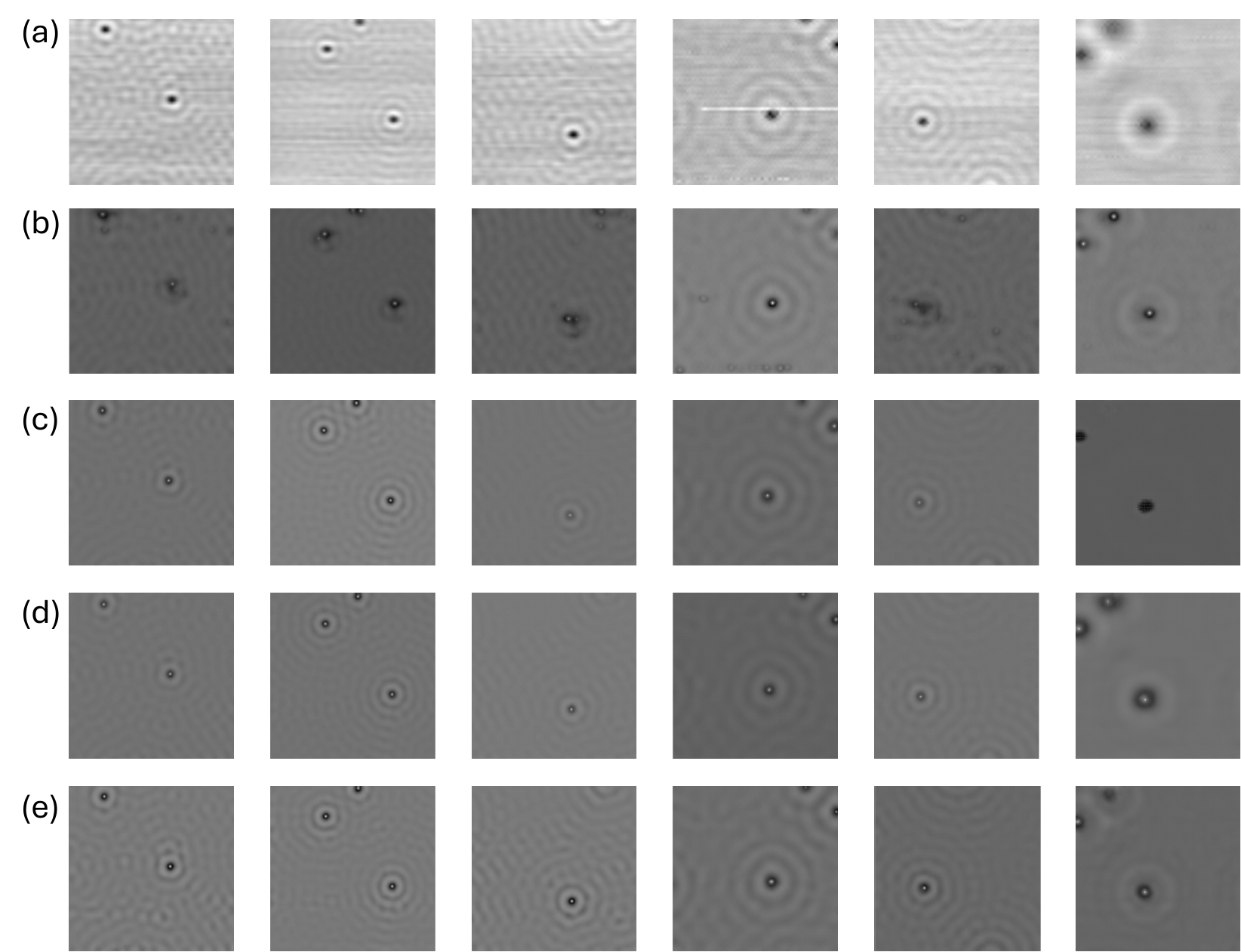}
	\caption{Examples of (a) Experimental images $I_E$; (b) images deblurred from $I_{E}$ using the pretrained CycleGAN model without domain adversarial (DA) module. (c) images deblurred from $I_{E}$ using CycleGAN and DA module; (d) images deblurred from $I_{E}$ using CycleGAN and DA with weight sharing; (e) images deblurred from $I_{E}$ using the proposed PDA-Net model, i.e., $I_{E}^D$. 
	% {\color{red}Model names should be consistent. Please put the results from our model in the last row. Please be consistent with Fig. 3.}
 }
	\label{Fig:lab_result}
\end{figure*}

    \subsection{Results for deblurring the simulated blurry images}
    \label{Subsec:Simulated}
    
The goal of the current investigation is to denoise the simulated blurry STM images shown in Fig. \ref{Fig:sim_result}(b), and compare them with the original simulated clear images presented in Fig. \ref{Fig:sim_result}(a). We conduct an ablation study to assess the impact of each module designed in our PDA-Net on the deblurring performance in the simulation domain. The denoising performance is benchmarked with the results produced by the sole cycle-consistency (i.e., CycleGAN) model without domain adaptation, weight sharing, or feature alignment. We present both quantitative and qualitative results for three variants of our methodology, with each variant incorporating additional components progressively: (1) cycle-consistency module enhanced with domain adversarial (DA) module (i.e., CycleGAN+DA), (2) cycle-consistency and DA module with shared weights (i.e., CycleGAN+DA+WS), and (3) cycle-consistency and DA module with both shared weights and feature alignment (i.e., PDA-Net).

The $MSE$, $PSNR$, and $SSIM$ for each variant and original CycleGAN model are presented in Table \ref{Table:sim}. The visual comparisons are shown in Fig. \ref{Fig:sim_result}(d,e,f), respectively. According to Table \ref{Table:sim}, CycleGAN generates a high $MSE$ and relatively low $PSNR$ with a high $SSIM$, along with the good deblurring results visualized in Fig. \ref{Fig:sim_result}(c). This indicates that while the pixel-wise error ($MSE$) and signal-to-noise ratio ($PSNR$) suggest room for improvement, the structural similarity ($SSIM$) and visual inspection confirm that the sole CycleGAN model can effectively preserve image details and structures when sufficient unpaired training data are provided. After adding the domain adversarial module (i.e., CycleGAN+DA), there is a notable reduction in $MSE$ and an increase in $PSNR$, albeit with a slight drop in $SSIM$. Incorporating weight sharing (i.e., CycleGAN+DA+WS) further improves $MSE$ and $PSNR$ while maintaining a high $SSIM$. Finally, the proposed PDA-Net model, which includes feature alignment, yields the best overall performance with the lowest $MSE$ of 513.06, highest $PSNR$ of 25.54, and highest $SSIM$ of 0.9332. 

%While initially, we hypothesized that the addition of domain adaptation upon the cycle-consistency module might introduce adverse effects due to potential feature interference, the results demonstrate otherwise. 

   \begin{table}
	\caption{The ablation study on denoising simulated blurry images.}
	\label{Table:sim}
	\vspace{-0.3cm}% Workaround to be conform with the .doc style. Only for table captions.
	\renewcommand\arraystretch{1.5}
	\setlength{\tabcolsep}{10pt} % Adjust the value to increase the space between columns
	\begin{center}
		\begin{tabular}{c| c c c}
			\hline
			& $MSE$ & $PSNR$ & $SSIM$ \\ \hline
			CycleGAN & 1243.33 & 23.14 & 0.9002 \\ \hline
			CycleGAN+DA & 1144.40 & 23.43 & 0.8840 \\ \hline
			CycleGAN+DA+WS & 822.54& 24.65 & 0.8938 \\ \hline
			PDA-Net  & 513.06 & 25.54 & 0.9332 \\ \hline
		\end{tabular}
	\end{center}
	\caption*{\footnotesize Note: DA refers to Domain Adversarial module, WS refers to Weight Sharing, and the proposed PDA-Net further includes feature alignment.}
\end{table}

  Compared with a pure CycleGAN, our PDA-Net achieves an improvement of 58.74\% on $MSE$, 10.37\% on $PSNR$, and 3.67\% on $SSIM$, demonstrating a comprehensive enhancement in both pixel-wise accuracy and structural similarity. The performance enhancement is due to the novel design of our PDA-Net. Specifically, domain adaptation allows the PDA-Net to better generalize across both simulated and real experimental domains, improving overall denoising performance. Weight sharing not only improve the training efficiency but also retains beneficial features learned from both generators (i.e., $G_D$ and $G_{DA}$), enabling effective knowledge sharing across both domains. Feature alignment ensures that the lower-level features extracted from both generators are harmonized, leading to a more robust and coherent representation of the image data across both domains.

\begin{table}
	\caption{The ablation study on denoising real-world experimental images. %{\color{red}Model names should be consistent.}
 }
	\label{Table: lab}
	\vspace{-0.3cm}% Workaround to be conform with the .doc style. Only for table captions.
	\renewcommand\arraystretch{1.5}
	\setlength{\tabcolsep}{10pt} % Adjust the value to increase the space between columns
	\begin{center}
		\begin{tabular}{c| c c}
			\hline
			& $BRISQUE$& $PIQE$  \\ \hline
			Experimental Images ($T_E$) & 69.96 & 96.15  \\ \hline
			CycleGAN & 114.40 & 59.14  \\ \hline
			CycleGAN+DA & 78.93 & 70.11  \\ \hline
			CycleGAN+DA+WS & 69.80 & 71.77  \\ \hline
			PDA-Net  & 52.99& 56.80 \\ \hline
		\end{tabular}
	\end{center}
%	\caption*{\footnotesize Note: DA refers to Domain Adaptation, WS refers to Weight Sharing, and the proposed PDA-Net further includes feature alignment.}
\end{table}

\subsection{Results for deblurring the real-world experimental images}
Obtaining noise-free STM images is impractical due to environmental and equipment limitations.  This limitation poses a significant challenge for existing denoising models, as they often rely on access to clear reference images for training and evaluation. Moreover, collecting a sufficient number of experimental STM images for unsupervised denoising is costly and time-consuming. Given these constraints, existing methods struggle to generalize effectively to real-world STM data. In this experimental study, we evaluate the proposed PDA-Net, which, to the best of our knowledge, is the first model specifically designed to operate under these constraints. As no directly comparable works exist in the literature, we benchmark our denoising results on experimental images against a CycleGAN model pretrained on the simulated dataset. Additionally, an ablation study is conducted to assess the effectiveness of key components, including domain adaptation, weight sharing, and feature alignment modules.

Fig. \ref{Fig:lab_result} provides visual examples of the denoising results in the real experimental domain. Fig. \ref{Fig:lab_result}(a) shows the experimental STM images of the Cu atom. %Fig. \ref{Fig:lab_result}(b-e) presents the deblurred images using different model variants. 
Fig. \ref{Fig:lab_result}(c) presents the denoising results from CycleGAN+DA, where noise reduction is achieved for most experimental images. However, some disadvantages still appear. For example, the last image in Fig. \ref{Fig:lab_result}(c) does not deblur to the expected denoised pattern. The absence of a central hollow/white dot in the middle of the atom indicates a non-ideal restoration of the atomic structure, and the electron wave pattern fails to appear as expected. Fig. \ref{Fig:lab_result}(d) depicts the results generated by CycleGAN+DA+WS, which display improved image clarity. Specifically, the details of the images, such as the wave pattern, start to become more clear, as seen in the third image in Fig. \ref{Fig:lab_result}(d). The center of the Cu atom also becomes clear as shown in the last image of Fig. \ref{Fig:lab_result}(d). However, the electron pattern is still not completely watched by the model. Fig. \ref{Fig:lab_result}(e) demonstrates the performance of the proposed PDA-Net. The wave patterns are fully captured by the model for all images, and the centers of Cu atoms are clearly presented. The proposed PDA-Net delivers the best denoising results with minimal artifacts and clear details. In contrast, Fig. \ref{Fig:lab_result}(b) presents the deblurring results of experimental images using the pretrained CycleGAN model learnd exclusively on the simulated dataset, without the DA module or WS and feature alignment techniques, which displays significant distortion compared to any variant of our PDA-Net.

Table 2 presents the quantitative results on real-world experimental STM images, evaluated using the $BRISQUE$ and $PIQE$. The baseline experimental images have a $BRISQUE$ score of 69.96 and a $PIQE$ score of 96.15. The CycleGAN model shows a degradation in perceived quality with a $BRISQUE$ score of 114.40, while the $PIQE$ score improves to 59.14. Adding the DA module (CycleGAN+DA) significantly reduces the $BRISQUE$ score to 78.93 but slightly increases the $PIQE$ score to 70.11. Incorporating WS (CycleGAN+DA+WS) further enhances the results, achieving a $BRISQUE$ score of 69.80 and a $PIQE$ score of 71.77. The proposed PDA-Net model, which further includes feature alignment, achieves the best results with a $BRISQUE$ score of 52.99 and a $PIQE$ score of 56.80, demonstrating superior image quality as perceived by both metrics.

%According to Table \ref{Table: lab}, the metric scores do not monotonically decrease, especially for $PIQE$, as more novel modules are added to the network architecture. The non-monotonic behavior of the metric scores can be attributed to the differing sensitivities of $BRISQUE$ and $PIQE$ to various types of image distortions and the specific characteristics of the denoising process. We figure that $BRISQUE$ is likely more reasonable than $PIQE$ because $BRISQUE$ relies on spatial domain features extracted directly from the image itself, which can better capture the unique textures and structural details inherent in STM images.%, allows itself to adapt to the specific characteristics of STM images more effectively. 
%On the other hand, $PIQE$ focuses on detecting perceptually significant distortions based on a block-wise analysis, where it divides the image into smaller blocks and evaluates the quality of each block separately. %, emphasizing local distortions. 
%This might not accurately reflect the subtle and complex features of STM images. Consequently, $BRISQUE$'s feature-based assessment provides a more suitable evaluation of STM image quality.

\section{Conclusions}
In this paper, we develop a novel framework, PDA-Net, for real-world STM image denoising. We first leverage physics principles to generate simulated clear/blurry STM images, which serves as the foundation to construct the denoising model for real-world experimental image denoising. %An auxiliary dataset of simulated blurry STM images is created to guarantee the denoising generator preserves the original image pattern. 
Second, we incorporate a cycle-consistency module to ensure the reliability of the denoising process in the simulation domain. Innovatively, we introduce a domain adversarial module to guide the domain adaptation generator in producing denoised STM images in the real experimental domain that are similar to the simulated clear benchmark. Moreover, feature alignment and weight sharing are engaged to further enable knowledge transfer from the simulated domain to the real-world denoising domain adaptation generator, thereby improving the performance of real-world denoising tasks. PDA-Net overcomes the limitations of traditional unsupervised denoising methods that typically require a set of clear images for training. Finally, we evaluate the proposed PDA-Net on both simulated and real-world STM datasets, with experimental results confirming its effectiveness in enhancing image quality. The PDA-Net framework not only facilitates more accurate analysis and interpretation in quantum material research but also has the potential for application in other complex imaging scenarios, such as medical image denoising.

\section{Acknowledgement}

  This research has been partially supported by a seed grant from the AI Tennessee Initiative at the University of Tennessee Knoxville. This research was partially supported by the National Science Foundation Materials Research Science and Engineering Center program through the UT Knoxville Center for Advanced Materials and Manufacturing (DMR-2309083). Author Jianxin Xie acknowledges the start-up funding support from the University of Virginia, School of Data Science. Author Bing Yao also acknowledges the funding support from the National Heart, Lung, And Blood Institute of the National Institutes of Health under Award Number R01HL172292.

\bibliographystyle{IEEEtran}

\bibliography{references}

\end{document}